# Parallel Minimum Spanning Tree Algorithms and Evaluation

Suryanarayana Murthy Durbhakula
Indian Institute of Technology Hyderabad
cs15resch11013@iith.ac.in, murthy.durbhakula@gmail.com

**Abstract.** Minimum Spanning Tree (MST) is an important graph algorithm that has wide ranging applications in the areas of computer networks, VLSI routing, wireless communications among others. Today virtually every computer is built out of multi-core processors. Hence it is important to take advantage of such parallel computing power by parallelizing existing algorithms and applications. Most of the earlier work on parallelizing MST focused on algorithms for PRAM models. There are two limitations to such studies. First, PRAM models assume infinite memory bandwidth which is unrealistic. Second, PRAM model based algorithms require at least O(n) processors where n being total number of vertices. For large graphs this is infeasible. There are very few implementations which target real systems. In this paper I present and evaluate two new parallel MST algorithms that are a variant of Parallel Boruvka algorithms: i) First algorithm uses lock variables without spin-locks ii) Second algorithm uses only atomic compare-and-swap (CAS) primitive. I evaluated the performance of these algorithms on a six-core, 12-thread Intel system on various input graphs of sizes up to 1 million vertices. First algorithm showed a speedup of up to1.94 over an un-optimized sequential algorithm and a speedup of up to 1.4 over an optimized sequential algorithm. Second algorithm showed a speedup of up to 2.03 over an un-optimized sequential algorithm and a speedup of up to 1.403 over an optimized sequential algorithm. When second algorithm using CAS is compared with the first algorithm second algorithm is found to be up to 1.15 times better than the first algorithm at four threads.

## 1 Introduction

Minimum spanning tree (MST) is an important graph algorithm. It is defined as a tree that connects all the vertices in a connected graph using edges such that the total sum of the edge weights in the tree is minimal. MST has wide ranging applications in the area of computer networks, VLSI routing, approximation algorithms for Travelling Salesman Problem, etc. There are many well-known sequential algorithms that compute MST of a graph. Some of them are:

1) Boruvka's Algorithm
2) Prim Algorithm
3) Kruskal's Algorithm

In this paper I present two new parallel algorithms for MST based on Boruvka's algorithm [1] and its implementation and evaluation on a 12-core Intel machine. In section 2, I first present the sequential Boruvka algorithm followed by parallel Boruvka algorithms. In section 3, I describe the methodology I used to evaluate algorithms. In section 4, I present results of parallel implementations for various graphs and show that these algorithms have speedup improvements till 8 threads. In section 5 I present related work and in section 6 I present conclusions and future work.

## 2 Parallel Minimum Spanning Tree Algorithms and Implementation

### 2.1 Sequential Boruvka Algorithm

Boruvka's algorithm works by initializing each vertex of the graph as a component by itself. It then finds for each component C1 a minimum edge E that connects it to another component C2. It adds such an edge E to the set of MST edges "M" and combines component C1 and C2 into a new combined component C3. Every component is identified by a unique root/parent vertex P. When combining two components C1 and C2 the algorithm assigns parent of the bigger component, among C1 and C2, as the parent vertex P of the combined component C3. Thus component C3 gets identified with vertex P. This process continues until all the components are combined into a single component. When the process completes the set "M" contains the list of MST edges and the tree formed by those edges is the MST of the graph. Below is the pseudo code for the sequential Boruvka's algorithm.

1. **Inputs:**
2. i) Input is a connected graph with edges having distinct weights
3. ii) Initialize a forest F of single vertex trees
4. **Data structures**:
5. i) Array of components: components[] with a root vertex associated with
6. each component.
7. ii) Array of minimum[] with size same as number of vertices. minimum[i] contains a pointer to minimum edge emanating from vertex i where vertex i is root of some component C.
8. iii) graph_edge is an array of edges where each edge E has attributes "src", "dest", and "weight" indicating source vertex of edge, destination vertex of edge, and weight of the edge respectively. graph is undirected hence src and dest are interchangeable and only named for readability. This applies to all the algorithms in this paper.

9. **Functions:**
10. i) int find(components[], vertex v): This function returns root vertex of the component to which input vertex v belongs
11. ii) void UnionOfComponents(component C1, component C2): This function combines two components C1 and C2 into a new component C3. The root vertex of C3 will be either root vertex of C1 or C2 whichever is a bigger component.
12. **Algorithm PseudoCode:**
13. While F has more than one component
14.   Initialize minimum[] to -1 for all vertices
15.   For all edges E in graph  /*Comment: for loop */
16.     component1 = find(components, source_vertex_of_E);
17.     component2 = find(components, dest_vertex_of_E);
18.     if(component1==component2)
19.       continue; /*Comment: continue to next iteration of for loop */
20.     else {
21.       if((minimum[component1]==-1) OR (graph_edge[minimum[component1]].weight > graph_edge[E].weight) {
22.         minimum[component1] = E
23.       }
24.       if((minimum[component2]==-1) OR (graph_edge[minimum[component2]].weight > graph_edge[E].weight) {
25.         minimum[component2] = E
26.       }
27.     }
28.   end for
29. 
30.   for each vertex v in Graph  /*Comment: for loop */
31.     if(minimum[v]!=-1) {
32.       component1 = find(components, graph_edge[minimum[v]].src);
33.       component2 = find(components, graph_edge[minimum[v]].dest);
34.       if(component1==component2)
35.         continue; /*Comment: continue to next iteration of for loop */
36.       else {
37.         Add edge minimum[v] to the set M of MST edges
38.         UnionOfComponents(component1, component2);
39.       }
40.   end for
41. end while
42.  Resultant set M has MST edges and the tree formed by those edges is a MST

**Sequential Boruvka Algorithm**

Complexity of Boruvka's algorithm is O(mlog(n)) where m – number of edges and n – number of vertices.

**Optimization over Sequential Boruvka Algorithm**

In this section I present an optimization of the sequential algorithm that involves eliminating unnecessary edges that need not be processed. With every graph edge E I associate an attribute called "covered" which is initialized to zero at the start of the algorithm. Once we find that an edge E is an internal edge of a component, which happens when variable component1 is same as component2, we set the attribute "covered" of edge E to 1. Next time when we consider an edge E for MST we check if the attribute "covered" is set to 1. If so we ignore the edge. We also set "covered" attribute of a MST edge to 1 once we add it to set of MST edges.

**2.2 Parallel Boruvka Algorithms**

**2.2.1 Parallel Algorithm using Lock variables**

In order to parallelize the above algorithm we let every thread run the same algorithm except that we guard the function UnionOfComponents with permissions. Before a thread tid combines two components C1 and C2 it checks whether variables lock_tid[C1] and lock_tid[C2] are set to -1. If so it tries to get permissions by setting lock_tid[C1] and lock_tid[C2] to tid. It then again checks whether it was successful in setting those variables. Here I assume that sequential consistency is maintained. Once it is successful it combines C1 and C2 by calling UnionOfComponents. It is important to do this because this function needs to be serialized. Consider for instance there are two edges E1 and E2. Let edge E1 be the minimum edge connecting components C1 and C2 and edge E2 be the minimum edge connecting component C2 to C3. If we do not serialize the UnionOfComponents then C2 can get absorbed into both component C1 and C3. And both edges E1 and E2 will get added as MST edges. This can lead to correctness issues. A given component should not be part of two different sibling components. Further, serial execution of above mentioned union of components can lead to different results compared to concurrent execution without lock protection. Consider, without any loss of generality, we first merge components C1 and C2. Lets us call this merged component C1-prime. Edge E1 gets added as an MST edge. Now there could be another minimum edge E3 from C1-prime connecting C1-Prime to C3. This edge E3 merges C1-Prime and C3 and gets added as MST edge instead of E2. Hence it is not correct to add edge E2 to MST in this case. Thus it is important to guard UnionOfComponents with permissions using lock_tid[] variables. Below is the pseudo-code of the parallel Boruvka algorithm

1. **Inputs:**
2. i) Input is a connected graph with edges having distinct weights. Initialize covered attribute of graph edge to 0.
3. ii) Initialize a forest F of single vertex trees
4. **Data structures**:
5. i) Array of components: components[] with a root vertex associated with

6. each component.
7. ii) Array of minimum[] with size same as number of vertices. minimum[i] contains a pointer to minimum edge emanating from vertex i where vertex i is root of some component C.
8. iii) graph_edge is an array of edges where each edge E has attributes "src", "dest", and "weight" indicating source vertex of edge, destination vertex of edge and weight of the edge respectively
9. iv) owner_tid[] is an array whose entry owner_tid[v] represents the thread which owns vertex v. Its size is same as total number of vertices in the graph.
10. v) lock_tid[] is an array whose entry lock_tid[v] represents the thread which owns permission to combine the component represented by vertex v with another component.

**11. Functions:**
12. i) int find(components[], vertex v): This function returns root vertex of the component to which input vertex v belongs
13. ii) void UnionOfComponents(component C1, component C2): In this function a thread tid combines component C2 into C1 and then sets owner_tid[C1] = tid.

**14. Algorithm PseudoCode for every thread with id "tid":**
15. While F has more than one component
16.   Every thread tid which owns vertex v, owner_tid[v]=tid, will initialize minimum[v] to -1 for all vertices it owns
17.   /* Comment: To reduce conflicts among threads start edge of every thread is spaced out. However every thread runs through all edges of the graph */
18.   for all edges E in graph
19.     if(graph_edge[E].covered==1) {
20.        continue;
21.     }
22.     component1 = find(components, source_vertex_of_E);
23.     component2 = find(components, dest_vertex_of_E);
24.     if(component1==component2) {
25.       graph_edge[E].covered = 1;
26.       continue;
27.       } else {
28.       if(owner_tid[component1]==tid) AND ((minimum[component1]==-1) OR (graph_edge[minimum[component1].weight > graph_edge[E].weight)) {
29.         minimum[component1] = E
30.       }
31.       if((owner_tid[component2]==tid) AND ((minimum[component2]==-1) OR (graph_edge[minimum[component2].weight > graph_edge[E].weight)) {
32.         minimum[component2] = E
33.       }
34.   }
35.   end for

36.
37.    For every vertex v in the graph
38.      if((owner_tid[v]==tid) AND (v is root vertex of a component) AND (minimum[v]!=-1) ){
39.          component1 = find(components, graph_edge[minimum[v]].src);
40.          component2 = find(components, graph_edge[minimum[v]].dest);
41.          if(component1==component2) {
42.             graph_edge[minimum[v]].covered=1;
43.             continue;
44.          } else {
45.
46.           if((lock_tid[component1]==-1) AND (lock_tid[component2]==-1)){
47.              lock_tid[component1]=tid;
48.              lock_tid[component2]=tid;
49.           }
50.
51.           if((lock_tid[component1]==tid) AND    (lock_tid[component2]==tid)) {
52.              component1_var = find(components, graph_edge[minimum[v]].src);
53.              component2_var = find(components, graph_edge[minimum[v]].dest);
54.
55.             if(component1_var!=component2_var) {
56.               Add edge minimum[v] to the set M of MST edges
57.               UnionOfComponents(component1_var, component2_var);
58.               graph_edge[minimum[v]].covered = 1;
59.             }
60.             lock_tid[component1]=-1;
61.             lock_tid[component2]=-1;
62.          }
63.    end for
64.  end while
65.   Resultant set M has MST edges and the tree formed by those edges is a MST

**Parallel Boruvka Algorithm using Lock variables**

   In the above algorithm at line 46 I check whether lock variables for component1 and component2 are available, that is set to -1. If so I set those variables to thread id tid so that I can combine both the components. At line 51 I check lock_tid variables again to make sure tid really has permission to combine both the components. At lines 52 and 53 I again call find to make sure both the components have not already been combined before combining them and adding the edge to MST. Also line 17 indicates a minor optimizations done to reduce collision among threads. While looping through all the edges of the graph every thread starts at a different evenly spaced out starting edge. For instance if there are 100 edges in the graph and we are parallelizing for four threads. Then thread0 starts at edge 0, thread1 starts at edge 25, thread2 starts at edge 50, thread3 starts at edge 75. However they cover all 100 edges. Threads 1, 2, and 3 will loop around and start at edge 0 once they reach edge 100 to cover all edges.

Below I present proof that the above algorithm does finds MST edges.

1) Every edge added in line 56 is a MST edge

This is true because minimum[v] is only calculated by thread tid which owns
vertex or component v; v is root vertex of the component. In the first for loop in lines 28 and 31 we compare minimum[v] with every graph edge and assign only that edge which is the minimum edge emanating from component v to minimum[v]. It is possible that while we are doing this component which owns v can get merged into a different component. However on line 38 we make sure this is not the case before proceeding. It is also possible that while thread tid is finding minimum in lines 28 and 31 another thread with id "tid_other" can merge some other component into the component which owns vertex v. However when that happens we set own_tid[v] to be the "tid_other".  Hence for tid on line 38 for vertex v owner_tid will no longer be tid. Due to this we will only process minimum edges in the for loop which starts at line 37. And before we add the edge to MST in line 56 we make sure the graph edge minimum[v], that is graph_edge[minimum[v]], still connects two different components. By the time we get permissions on component1 and component2 both components could have been merged into a single component. That's why we check again on line 55. And it is also possible that component1 could have been merged into component1_var and component2 could have been merged into another component component2_var. Still minimum[v] is a minimum edge that connects these two components. To prove this lets call component1_var as C3 and component2_var as C4. Let's call component1 as C1 and component2 as C2. Below is a diagram which shows connection between these vertices.

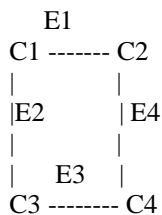

```
     E1
C1 ------- C2
|          |
|E2        | E4
|          |
|   E3     |
C3 -------- C4
```

E1 is minimum of all edges which connects C1 to C2. Thus E1 is edge represented by minimum[v] from the description above. E2 is minimum of all edges
that connect C1 to C3. E3 be minimum of all edges that connect C3 to C4
and E4 be minimum of all edges that connect C4 to C2.

Without any loss of generality let's assume E1 is the minimum edge emanating from C1. E2 is the minimum edge emanating from C3 and E4 is the minimum edge emanating from C4. We make following observations:

1) E1 is less than E2 otherwise E2 will also become minimum edge emanating from C1.
2) E2 is less than E3 otherwise E3 will become the minimum edge emanating from C3.
3) E4 is less than E3 otherwise E3 will become minimum edge emanating from C4.

Hence from above we conclude that E1 is less than E3. When C3 combines component C1 into itself E2 becomes an internal edge and when C4 combines component C2 into itself E4 becomes an internal edge. Thus edges E1 and E3 now are edges that run between components C3 and C4. Among them E1 is less than E3 as I have shown from observations above. Thus E1 is part of MST. In other words minimum[v] on line 56 is part of the MST.

2) The edge that is added in line 56 will not create a cycle in the set of MST edges

Before adding the edge to MST in line 56 we get permission to access components and we also make sure that the edge with id minimum[v] still connects two different components after we get permissions. An edge that forms a cycle will not be connecting two different components. Hence the MST edge added in line 56 will not form a cycle.

### 2.2.2 Parallel Algorithm using Atomic Compare and Swap variable

The structure of this algorithm and its proof is very similar to the algorithm which uses lock variables. Hence I am not again giving the proof here. Its pseudo code is given below. There are two primary differences compared to the previous algorithm. i) There are no lock variables. ii) When combining two components C1 and C2 in line 45 below I used atomic compare and swap instruction. Using this construct I atomically set parent attribute of component C2 to be C1. Hence all the children of component C2 will become children of component C1 thus merging component C2 to C1.

1. **Inputs:**
2. i) Input is a connected graph with edges having distinct weights. Initialize covered attribute of graph edge to 0.
3. ii) Initialize a forest F of single vertex trees
4. **Data structures**:
5. i) Array of components: components[] with a root vertex associated with
6. each component.
7. ii) Array of minimum[] with size same as number of vertices. minimum[i] contains a pointer to minimum edge emanating from vertex i where vertex i is root of some component C.

8. iii) graph_edge is an array of edges where each edge E has attributes "src", "dest", and "weight" indicating source vertex of edge, destination vertex of edge and weight of the edge respectively
9. iv) owner_tid[] is an array whose entry owner_tid[v] represents the thread which owns vertex v. Its size is same as total number of vertices in the graph.
10. **Functions:**
11. i) int find(components[], vertex v): This function returns root vertex of the component to which input vertex v belongs
12. ii) void UnionOfComponents(component C1, component C2): In this function a thread tid combines component C2 into C1 and then sets owner_tid[C1] = tid.
13. **Algorithm PseudoCode for every thread with id "tid":**
14. While F has more than one component
15.   Every thread tid which owns vertex v, owner_tid[v]=tid, will initialize minimum[v] to -1 for all vertices it owns
16.   /* Comment: To reduce conflicts among threads start edge of every thread is spaced out. However every thread runs through all edges of the graph */
17.   for all edges E in graph
18.     if(graph_edge[E].covered==1) {
19.       continue;
20.     }
21.     component1 = find(components, source_vertex_of_E);
22.     component2 = find(components, dest_vertex_of_E);
23.     if(component1==component2) {
24.       graph_edge[E].covered = 1;
25.       continue;
26.     } else {
27.       if(owner_tid[component1]==tid) AND ((minimum[component1]==-1) OR (graph_edge[minimum[component1].weight > graph_edge[E].weight)) {
28.         minimum[component1] = E
29.       }
30.       if((owner_tid[component2]==tid) AND ((minimum[component2]==-1) OR (graph_edge[minimum[component2].weight > graph_edge[E].weight)) {
31.         minimum[component2] = E
32.       }
33.     }
34.   end for
35.
36.   For every vertex v in the graph
37.     if((owner_tid[v]==tid) AND (v is root vertex of a component) AND (minimum[v]!=-1) ){
38.       component1 = find(components, graph_edge[minimum[v]].src);
39.       component2 = find(components, graph_edge[minimum[v]].dest);
40.       if(component1==component2) {

```
41.         graph_edge[minimum[v]].covered=1;
42.         continue;
43.      } else {
44.         /*Comment: Below function is a one line function which
   atomically sets parent of component2 to be component1 */
45.         UnionOfComponents(component1, component2);
46.      if(CAS successful) {
47.         owner_tid[component1] = tid;
48.         Add edge minimum[v] to the set M of MST edges
49.         graph_edge[minimum[v]].covered = 1;
50.      }
51.   end for
52. end while
53.   Resultant set M has MST edges and the tree formed by those edges is a MST
```

**Parallel Boruvka Algorithm using atomic CompareAndSwap**

In the next section I describe the methodology I used to evaluate the algorithm.

## 3 Methodology

I have implemented both sequential and parallel Boruvka's algorithms in C++. I have also written a graph generator program that generates graphs of various sizes. It takes size of the graph and an average vertex degree of the graph as inputs and generates various graphs. I have evaluated the parallel implementation using these graphs of sizes starting from 10000 nodes all the way up to 1 million nodes and varied the average vertex degree from 3 to 6 to 9. Below in table 1 I summarize the input graphs used in this study.

**Table 1: Input graphs and their description**

| Graph Name | Description |
|---|---|
| Graph10K_3 | 10K vertex graph with average vertex degree 3 |
| Graph10K_6 | 10K vertex graph with average vertex degree 6 |
| Graph10K_9 | 10K vertex graph with average vertex degree 9 |
| Graph100K_3 | 100K vertex graph with average vertex degree 3 |

| Graph100K_6 | 100K vertex graph with average vertex degree 6 |
|---|---|
| Graph100K_9 | 100K vertex graph with average vertex degree 9 |
| Graph1M_3 | 1 million vertex graph with average vertex degree 3 |
| Graph1M_6 | 1 million vertex graph with average vertex degree 6 |
| Graph1M_9 | 1 million vertex graph with average vertex degree 9 |

Below in table2 I describe the configuration of the Intel system I have used for evaluating the parallel MST. In the next section I describe results of the evaluation.

**Table 2: Configuration Parameters**

| Parameter | Value |
|---|---|
| System | Intel Xeon E-2186G |
| CPU Frequency | 3.86 GHz |
| Number of cores | 6 |
| Number of threads | 12 |
| Cache size | 12MB |
| DRAM size | 31GB |

## 4 Results

In this section I first present performance improvement of the sequential algorithm improvement described in section 2.1. Figure 1 presents those results. We can see that the performance improvement varies between 24% and 45%. Figure 2 and 3 presents performance improvement results of the two parallel Boruvka algorithms as we vary the number of threads from 1 to 16 threads. All the numbers in the graph are in milliseconds. In Figure 2 we can see that maximum speed-up for the algorithm using lock variables is 1.94 over un-optimized sequential algorithm and 1.4 over optimized sequential algorithm and it occurs at four threads. . Similarly in Figure 3 we can see that maximum speed-up for the algorithm using atomic CAS is 2.03 over un-optimized sequential algorithm and 1.403 over optimized sequential algorithm and it occurs at four threads. From both figure 2 and 3 we can see from that performance improves as we increase the number of threads from 1 to 2 to 4 and sometimes up to eight threads. However when we increase the number of threads to 16 there is a drop in performance. This is because the machine on which the algorithm is running can only support a maximum of 12 hardware threads. At 16 threads the number of software threads exceeds the number of hardware threads thus degrading the performance. This is consistently seen across all problem sizes. In some cases performance also drops at eight threads because the Intel system actually has six cores with each core supporting execution of up to two threads using hyper-threading. Sometimes hyper-threading does not lead to performance benefits giving rise to slowdown at eight threads. Figure 4 shows performance improvement of the algorithm using atomic CAS over the algorithm using lock variables at four threads. We can see that performance improves up to 15.5% using atomic CAS.

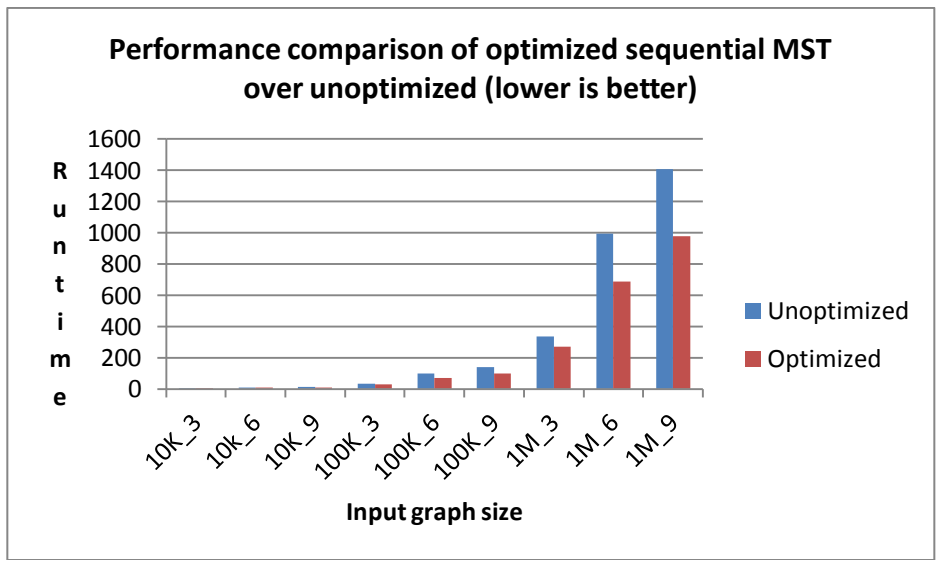

**Figure 1: Performance comparison of optimized sequential MST over unoptimized**

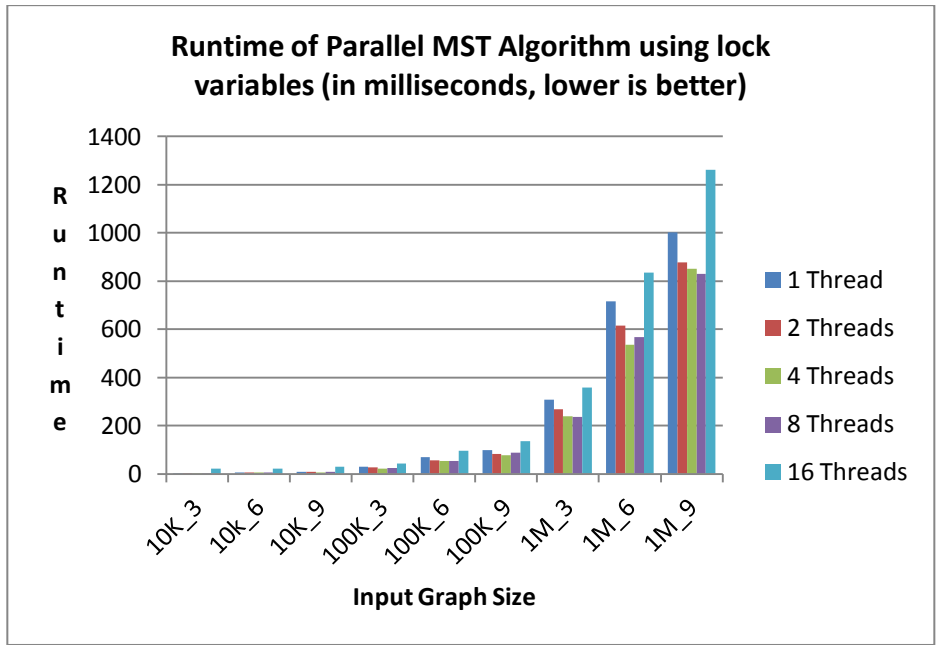

Figure 2: Performance improvement of Parallel MST Algorithm using lock variables

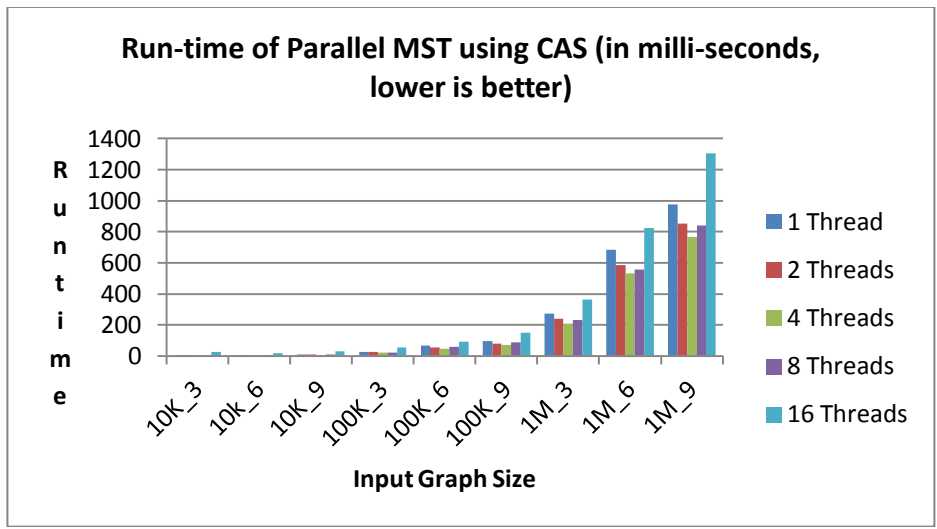

Figure 3: Performance improvement of Parallel MST Algorithm using atomic CAS

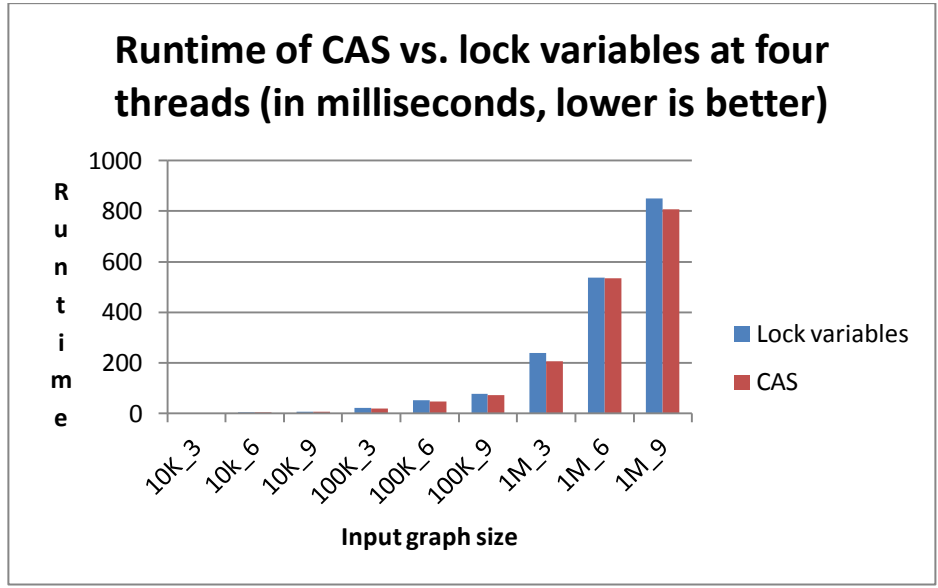

**Figure 4: Performance improvement of Parallel MST Algorithm using atomic CAS over Parallel MST Algorithm using lock variables**

## 5 Related Work

Vishkin [2] proposed algorithms for CRCW PRAM model. Similarly, Johnson and Metaxas [3] proposed algorithms for EREW PRAM model. For large graphs, both of the above algorithms are impractical as they would require at least O(n) processors, where n is number of vertices. There are earlier studies on parallel implementation of Boruvka's algorithm. Chung et al [4] studied an implementation of parallel Boruvka's algorithm on CM-5 machine using message-passing for graphs up to 64000 vertices. They paper a speed-up of a factor of 4 on 16 processors. Bader et al [5] use a hybrid approach of combining Boruvka and Prim algorithm and paper a speedup of 4 on 8 processors on a Sun E4500 machine. Setia et al [6] proposed a new parallel Prim algorithm and reported a best case speedup of up to 2.64 on four threads. However the size of the graphs they studied are very small ranging from 1000 to 5000 nodes. Whereas the sizes of the graphs we studied are large and realistic, up to a million nodes, and they would not fit in the cache.

## 6 Conclusions and Future work

MST is an important graph algorithm that has wide ranging applications in computer networks, VLSI routing, approximation algorithms for TSP among others. In this paper I have presented two parallel algorithms based on Boruvka's algorithm. First algorithm uses lock variables without spin locks. Second algorithm uses only atomic CAS construct. I have implemented both these algorithms and evaluated its performance on a 12-core Intel system. First algorithm showed a speedup of up to1.94 over an un-optimized sequential algorithm and a speedup of up to 1.4 over an optimized sequential algorithm. Second algorithm showed a speedup of up to 2.03 over an un-optimized sequential algorithm and a speedup of up to 1.403 over an optimized sequential algorithm. When second algorithm using CAS is compared with the first algorithm, second algorithm is found to be up to 1.15 times better than first algorithm at four threads.